# Plasmons dispersion and nonvertical interband transitions in single-crystal $Bi_2Se_3$ investigated by electron energy loss spectroscopy

S. C. Liou,[1] M.-W. Chu,[1] R. Sankar,[1] F.-T. Huang,[1] G.-J. Shu,[1] F. C. Chou,[1] and C. H. Chen[1,2*]

[1]*Center for Condensed Matter Sciences, National Taiwan University, Taipei 10617, Taiwan*

[2]*Department of Physics, National Taiwan University, Taipei 10617, Taiwan*

[*]Corresponding author: S. C. Liou (scliou0903@ntu.edu.tw)

**ABSTRACT**

Plasmons dispersion and nonvertical interband transitions in $Bi_2Se_3$ single crystals were investigated by electron energy-loss spectroscopy in conjunction with (scanning) transmission electron microscopy, (S)TEM-EELS. Both volume plasmons ($\pi$ plasmon at 7 eV and $\pi+\sigma$ plasmon at 17 eV) and surface plasmons (~5.5 and 10 eV) were demonstrated in STEM-EELS spectra and the corresponding spectral imaging in real space. In the further EELS experiments in reciprocal space, the momentum-dependent spectra reveal very different dispersion behavior between $\pi$ and $\pi+\sigma$ plasmons, with $\pi+\sigma$ plasmons showing a typical quadratic dependence whereas the $\pi$ plasmon exhibiting a linear dispersion analogous to what was reported for graphene. Furthermore, a low energy excitation 0.7~1.6 eV was also observed which is attributed to direct nonvertical interband transitions along the $\Gamma F$ direction.

# I. INTRODUCTION

Bismuth selenide (Bi$_2$Se$_3$) is a well-known old material with new discovery in condensed-matter physics. Bi$_2$Se$_3$ has initially been demonstrated to be a thermoelectric material with high figure of merit (ZT)[1-2] for the applications of thermoelectric and infrared devices.[3-4] In the past few years, Bi$_2$Se$_3$ has attracted tremendous attention in the field of topological insulator (TI) because it was a second-generation three-dimensional (3D) TI featuring single Dirac cone at the Γ point with the gapless surface state (or conducting surface state).[5-7] Furthermore, due to the strong spin-orbit coupling, Bi$_2$Se$_3$ posses the gapless surface state (or conducting surface state) and retains a bulk energy gap ($E_g \sim 0.3$ eV), opening a new avenue to applications ranging from spintronics to quantum computation.[5-7] Consequently, most research activities focus on the studies of conducting surface state and related surface band structure of Bi$_2$Se$_3$ using angle-resolved photoemission spectroscopy (ARPES) and first-principles calculations,[5-9] with only few investigations on the intrinsic bulk band structure of Bi$_2$Se$_3$.[10-13]

Structurally, Bi$_2$Se$_3$ consists of alternating layers of Bi and Se that are arranged along the *c* axis in five covalent bonded atomic layers with the sequence of Se–Bi–Se–Bi–Se forming a tightly bound quintuple layer (QL), and the QLs are weakly bounded with a van der Waals force.[14,15] This structure clearly reveals highly anisotropic bonding between the lateral (within QLs) and the longitudinal (between the QLs) directions. It would be quite interesting to study the electronic anisotropy in this layered material, similar to what has been observed in graphite.[16,17]

In contrast to photoemission spectroscopy which reveals only the occupied states of valence band, the electron energy-loss spectroscopy (EELS) can provide the information of both valence and conduction bands.[16] The low-loss region (energy loss < 50 eV) in EELS spectra contains the excitations of valence electrons including collective modes, e.g. surface and volume plasmons, and single-particle excitations, e.g. interband transitions, and low-lying core-level ionizations.[16,18] Furthermore, momentum-dependent EELS spectra can be obtained using a parallel beam in transmission electron microscopy (TEM), which reveals the dispersion of plasmons and other spectral excitations of the materials.[16-21] Electronic excitations of Bi$_2$Se$_3$ has been studied using EELS, revealing a volume plasmon at 16.5~ 17 eV, whereas the physical origin of a spectral feature at 7 eV was not clearly identified.[11] Furthermore, no momentum (*q*)-resolved EELS on the study of electronic excitations in $Bi_2Se_3$ has been reported.

Here, the electronic excitations in single-crystal $Bi_2Se_3$ were investigated using EELS in conjunction with (scanning) transmission electron microscopy, (S)TEM-EELS. The results of positions-dependent STEM-EELS spectra and real-space spectral imaging by STEM-EELS unambiguously demonstrated the presence of surface plasmons (spectral features at ~5.5 and 10 eV) at the sample edge with distinct evanescent wave fields extended into vacuum, and the volume plasmons (spectral features at 7 and 17 eV) in the material interior. The investigations of momentum-dependent EELS spectra revealed a linear and parabolic dispersion for the 7 and 17 eV volume plasmon, respectively. Moreover, a low energy excitation 0.7~1.6 eV was also observed which is attributed to the direct nonvertical interband transitions along the ΓF direction

## II. EXPERIMENTAL

The high purity bismuth (99.999%) and selenium (99.999%) powders with Bi:Se molar ratios of Bi : Se = 2 : 3 were mixed thoroughly in an argon-filled glove box. Single-crystal $Bi_2Se_3$ was further grown using the vertical Bridgman method. The detail procedure of growth process and parameters was reported in our previous study.[15] Two types of TEM specimens with the crystalline c-axis perpendicular and parallel to the sample plane were prepared for the momentum-dependent (*q*-dependent) EELS investigations. The sample with perpendicular c-axis was easily prepared by mechanical cleavage, whereas the sample with parallel c-axis was prepared by standard cross-sectional technique followed by $Ar^+$ ion milling at 3 keV (Gatan PIPS). Both specimens were finally cleaned by low-energy ion milling at 0.3 keV to remove the possible surface amorphous layers. All (S)TEM-EELS spectra were acquired on a FEI field-emission STEM/TEM, Tecnai F20, operated at 200 kV. Throughout the (S)TEM-EELS experiments, the electron monochromator was employed to improve the energy resolution from 0.6 eV to 0.24 eV. The momentum -dependent EELS experiments were carried out under diffraction mode with parallel beam illumination, and the beam-convergence semi-angle was 0.09 and 0.26 mrad (translating into momentum resolution of 0.06 $Å^{-1}$ and 0.23 $Å^{-1}$) without and with electron monochromator, respectively. The single scattering STEM-EELS spectra deconvoluted from raw data, subsequent Kramers-Krönig analysis (KKA) and the calculation of effective electron number ($n_{eff}$) were conducted on DigitalMicrograph EELS package, which was written on the basis of Ref. 18.

## III. RESULTS AND DISCUSSIONS

The EELS spectra obtained in STEM mode, STEM-EELS, provides the characterization of electronic excitations as a function of the electron-probe position on materials. Figure 1(a) shows the STEM–EELS spectra acquired on single-crystal $Bi_2Se_3$ with the incident electron probe traversed from the interior toward the edge of the sample in vacuum by a 2-nm step increment (color circles in high-angle annular dark field (HAADF) image, inset in Fig. 1(a)). The EELS spectra, normalized to the intensity of zero-loss peak (ZLP) and followed by the removal of ZLP with Fourier-log method, are vertically shifted for clarity. The pink curve indicated by the solid black arrow (Fig. 1(a)) exhibits the spectrum taken in vacuum and right at the sample edge. It should be noted that the convergence semi-angle of electron probe in monochromator-STEM mode is 13 mrad in this study. Thus, the experimental EELS spectrum in Fig. 1(a) integrates electronic contributions over a wide range of reciprocal-lattice vectors ($q$~ 3 Å$^{-1}$), and averages out all possible anisotropic electronic contributions.

We first discuss the EELS spectra when the electron probe is positioned well in the material interior. In this circumstance, several spectral features at ~7, ~17, ~26.4 and ~28.4 eV were clearly observed (black curve in Fig. 1(a)), consistent with the previously known results.[11] The two peaks at ~26.4 and ~28.4 eV can be identified as the Bi $O_{4,5}$ edges excited from Bi $5d$ electrons.[11] The predominant spectral feature at ~17 eV is characterized as the volume-plasmon in $Bi_2Se_3$, whereas spectral feature at ~ 7 eV was interpreted as the surface-plasmon.[11] In our studies, we found the intensity of the spectral features at 7 and 17 eV would increase with increasing sample thickness indicating that both spectral features at 7 and 17 eV were volume excitations. Consequently, the previous interpretation of the spectral feature at ~7 eV as surface-plasmon appears to be incorrect. More detailed discussion of physical nature of these two excitations will be given in later paragraphs. Further inspection of the same spectrum in Fig. 1(a) also reveals a broad feature from ~10 to ~12 eV, which was previously attributed to either interband transition (arising from Bi 6*s* electrons)[11] or collective excitations[13].

Considering that the EELS features can be generally understood in the framework of the macroscopic dielectric-response theory,[16,18,22] we thus derived the frequency ($\omega$)-dependent dielectric function, $\varepsilon(\omega)$, of $Bi_2Se_3$ from the black spectrum in Fig. 1(a) by performing Kramers-Krönig analysis (KKA), as shown in Fig. 1(b). Indeed, the $\varepsilon(\omega)$ we obtained from EELS spectrum not only matches well with previous results obtained from optical measurement in the energy loss range between 2.5 and 12 eV,[23] it also extends the spectral range up to 40 eV. We note, however, $\varepsilon(\omega)$ obtained in this study does not yield accurate results for the energy loss below 2.5 eV due to the ZLP tail, which muddles the weak features in this spectral regime. Hence, $\varepsilon(\omega)$ for the energy loss < 2.5 eV in Fig. 1(b) is actually reproduced from the optical measurement[23] to complete the integrity of $\varepsilon(\omega)$ in the low energy-loss regime. In addition to the $\varepsilon(\omega)$ derived from STEM-EELS spectra, we also derived $\varepsilon(\omega)$ from the TEM-EELS spectra and obtained similar results.

In Fig. 1(b), $\varepsilon_1$ passes through zero at ~ 7 and ~15.9 eV accompanied by a decreasing $\varepsilon_2$. This gives rise to the maxima in the volume loss function ($\propto \mathrm{Im}\left\{\dfrac{-1}{\varepsilon(\omega)}\right\}$) at the energies of ~ 7 and 17 eV (black curve, Fig. 1(a)), characteristic of volume-plasmon excitations.

Since the plasmon resonance originates from the collective oscillations of valence electrons in the crystal, the effective number $n_{eff}(\omega)$ of electrons per atom in $Bi_2Se_3$ can be calculated by using sum rules (inset, Fig. 1(b)). Unlike graphite, which is known to exhibit two distinct plateaus at 7 eV with $n_{eff}$~ 1 electron per atom and $n_{eff}$~ 4 electron per atom at 27 eV, signifying the clear distinction of $\pi$ and $\pi+\sigma$ plasmon excitation, respectively,[24] no obvious plateaus are visible in $Bi_2Se_3$. In order to have assurance of the technique (sum rules) we used, we have also reproduced the double plateau structure in graphite. Nevertheless, the effective number $n_{eff}(\omega)$ of electrons per atom (inset, Fig. 1(b)) does show discernible changes of slopes near 7 eV with $n_{eff}$ ~ 1 electron per atom and near 15 eV with $n_{eff}$~ 3 electrons per atom. Although the oscillator strength of $\pi$ and $\sigma$ electrons are not clearly defined in the present case, the spectral features at 7 eV and 17 eV can still be largely interpreted as a plasma oscillation of the $\pi$ and $\pi+\sigma$ electrons, respectively, similar to layer structure of graphite and molybdenite ($MoS_2$).[24,25]

Further inspection of the absorption features ($\varepsilon_2$) in Fig. 1(b) also reveals a rather weak and broad peak spanning from ~8 eV to ~ 12 eV, which occurs as a result of the interband-transition excitations (e.g. contribution of Bi 6*s* electrons).[11]

As the electron probe continues to move from the material interior (black curve in Fig. 1(a)) to the point near the edge (pink curve in Fig. 1(a)), the spectral feature at ~7 eV (volume plasmon) gradually red-shifts to ~5.5 eV ($\varepsilon$= -1.72 + $i$2.93, Fig. 1(b)), whereas the spectral feature at ~17 eV slightly red-shifts to ~16.5 eV with rapidly diminishing intensity. The intensity of interband transitions at 8~12 eV also decreases and becomes a broad peak at ~10 eV ($\varepsilon$= -0.98 + $i$2.14, Fig. 1(b)). Considering the negative $\varepsilon_1$ in the corresponding energy regime in Fig. 1(b), the spectral features at ~5.5 and ~10 eV indicate the presence of surface plasmons (SPs).[16,18] When the electron probe moves further away from the sample edge into vacuum (e.g. sky blue, blue and green, Fig. 1(a) and (b)), the spectral features at ~5.5 and ~10 eV are further weaken but still visible, signifying the characteristic of the evanescent wave fields of SPs.[16,18,22,26]

The STEM-EELS spectrum imaging,[27] which has been well established to probe SPs in noble metals[28-30] and semiconductor[31], was performed to further confirm the physical origin of the spectral features at ~5.5 and ~10 eV as SPs. Figure 1 (c)–(e) show the STEM-EELS intensity maps, measured from the same area in HAADF image (inset, Fig. 1(a)), of the spectral features at ~5.5, ~10 and 17 eV, respectively. The maximum intensity of the spectral features at ~5.5 eV and 10 eV (Fig. 1(c) and (d)) is clearly localized at the sample edge of the $Bi_2Se_3$ with evanescent wave field decaying into vacuum, characteristic of SPs excitations. In contrast, the maximum intensity of the spectral feature at 17 eV (volume plasmon) is strongly localized within the bulk interior (Fig. 1(e)) as expected. Combining the experimental results of STEM-EELS spectra and real-space spectral imaging, we conclude that the spectral features at ~5 and ~10 eV are indeed SPs.

As mentioned previously, possible electronic anisotropy could not be resolved in the STEM-EELS experiments due to the large convergence semi-angle of electron probe. To overcome this difficulty and study the electronic excitations in greater details, we also carry out the momentum-dependent EELS experiments to study the dispersion of spectral excitations under the parallel-beam electron diffraction mode in conventional TEM.

Figure 2 (a)-(c) show the momentum-dependent EELS spectra obtained along the ΓM, ΓK and ΓZ directions, respectively, for both volume plasmons excitations at 7 and 17 eV. Here we use the two-dimensional notation of the Brillouin zone since the significance of the three-dimensional Brillouin zone can be safely ignored for the collective excitations we are going to discuss below. The magnitude of ΓM, ΓK and ΓZ is 1.52 Å$^{-1}$, 1.75 Å$^{-1}$ and 0.68 Å$^{-1}$, respectively. The $q$-dependent EELS spectra along both ΓM and ΓK directions (Fig. 2(a) and (b)) clearly show that both π (7 eV) and π+σ plasmons (17 eV) peaks shift to higher energies with increasing $q$ values (dot arrows). It should also be noted that as the plasmon peaks disperse upward to higher energies with larger $q$, their peaks intensities decrease rapidly and become broader, and, beyond some point, plasmon excitations become poorly defined. For the π (7 eV) and π+σ plasmons (17 eV) peaks, the intensity decreases by three orders of magnitude with momentum transfer ($q$) changing from 0 to ~1.3 Å$^{-1}$. For example, π plasmon can be clearly identified for $q$ < 0.84 Å$^{-1}$, whereas it becomes a broad and poorly defined feature $q$ > 0.84 Å$^{-1}$.

Moreover, the presence of multiple scattering involving plasmons and phonons which often gives rise to an additional non-dispersive peak at large $q$ could further exacerbate the difficulty of analyzing plasmon dispersion at large $q$.[19] This occurrence can be clearly seen in Fig. 2(a) where π+σ plasmon becomes broad at the $q$ > 0.63 Å$^{-1}$ and splits into two peaks. The lower-energy one is non-dispersive (marked with dash line, Fig. 2(a)) due to multiple scattering and the other dispersive peak (dotted arrow), which clearly shifts to higher energy (up to ~23.5 eV) with increasing $q$, represents the true plasmon peak. It is difficult in the present case to avoid the multiple phonon scattering due to the large atomic number of Bi despite the small sample thickness of 0.3λ (λ is inelastic mean free path) used for $q$-dependent experiments in Fig. 2.

The $q$-dependent EELS spectra along ΓK direction shown in Fig. 2(b) exhibit great resemblance to that along ΓM. In contrast, the $q$-dependent EELS spectra along ΓZ direction (Fig. 2(c)) displays less dispersive plasmon peaks which shift slightly less than 1 eV to higher energy at $q$= 0.42 Å$^{-1}$. At larger $q$, π and π+σ plasmons peaks shift back to the original energies observed at $q$= 0, most likely due to the multiple scattering discussed above. A broad hump at ~19 eV is noticeable in the EELS spectrum for $q$= 0.42 Å$^{-1}$, which could represents the heavily damped π+σ plasmon peak near the zone boundary of ΓZ direction.

To accurately determine the plasmon dispersion, we have applied Lorentz fitting (gray and red curves in Fig. 3(a)) to the EELS spectra and obtained the precise spectral positions of each spectral features including both π, π+σ plasmons, and Bi $O_{4,5}$ edges. The energies of π and π+σ plasmons along ΓM, ΓK and ΓZ directions as a function as momentum transfer ($q$) are then plotted in Fig. 3(b) with black squares, gray open circles and red triangles, respectively. First of all, we note the great similarities between ΓM and ΓK in-plane directions, indicating a very weak in-plane anisotropy. From Fig. 3(b), a striking difference emerges for the dispersion between π and π+σ plasmons. While the π+σ plasmon displays the typical parabolic dispersion with $q^2$ dependence, the π plasmon reveals a linear dispersion as a function of momentum-transfer ($q$), similar to what was reported for the 2-dimensional plasmon in graphene.[32,33]

The plasmon dispersion relation for a 3-dimensional system is often described by the following equation:[16,18]

$$E_{plasmon}(q) = E_{plasmon}(0) + \alpha \frac{\hbar^2}{m_o} q^2$$

where $E_{plasmon}(q)$ is the $q$-dependent plasmon energy, $\alpha$ is the dispersion coefficient, and $m_o$ is effective electron mass. The π+σ plasmons dispersion shown in Fig. 3(b) can be well fitted by the above equation and the dispersion coefficient (α) is found to be 0.57 (blue dash curve, Fig. 3(b)). The plasmon dispersion along ΓZ is largely similar to the in-plane directions, despite its small Brillouin zone.

It is tantalizing to consider the possibility of 2D characteristic for the π plasmon in $Bi_2Se_3$ which shows a linear dispersion in $q$ similar to graphene.[32,33] The weak Van der Waals forces arisen from the π bonding between the two outmost Se[(1)] layers in the tightly bonded Se[(1)]–Bi–Se[(2)]–Bi–Se[(1)] quintuple layers (QL) may suggest the formation of two-dimensional electron gas. Recent research, in fact, has also proposed the concept of two-dimensional electron gas (2DEG) in $Bi_2Se_3$.[34] However, the lack of plateau in the $n_{eff}$ curve shown in Fig. 1(c) seems to indicate a significant mixture of π and σ character of the electronic structure and diminish the possibility of 2D electron gas formation of π electrons. It is evident that more work will be needed to clarify the issue of 2D electronic character of π electrons.

Finally, we would like to point out that a conspicuous spectral feature at low energy around 1 eV was observed in the $q$-dependent EELS spectra along the ΓF direction, shown in Fig. 4(a). A broad shoulder at ~0.7 eV appears for $q$= 0.2 Å$^{-1}$, becomes a pronounced spectral feature at 0.8~1.6 eV in the $q$ range of 0.2~ 0.7 Å$^{-1}$, and finally disappears at $q$=0.8 Å$^{-1}$. It should be noted that a sample tilt from the normal incidence is necessary for the momentum-transfer to reach the ΓF direction (ΓF would project into ΓM in two-dimensional considerations). This low energy peak is only weakly excited along the ΓM direction without sample tilting.

The dispersion of this low energy peak is plotted in Fig. 4(b) which reveals a linear dispersion and an energy of 0.35 eV when extrapolated to $q$=0. This energy coincides with the bandgap in $Bi_2Se_3$ ($E_g$=0.32 eV) within the accuracy of our experiments and, hence, it strongly suggests that the strongly dispersive peak shown in Fig. 4(a) is due non-vertical interband transitions. The linear dispersion is analogous to the direct nonvertical interband transition observed in aluminum.[18] The observed linear dispersion is found to follow closely to the conduction band dispersion along the ΓF direction,[12,35] also shown in Fig. 4(b). It appears that the non-vertical interband transition is characterized by an initial state at the Γ point of the valence band and the final states in the lowest conduction band in the ΓF direction. Despite the excellent agreement between the measured dispersion and the lowest conduction band along the ΓF direction, this comparison is most likely an overly simplified picture for the dispersion of non-vertical interband transitions. A complete calculation of both $q$ and $\omega$-dependent dielectric function is required to thoroughly understand the details of this low energy excitation. We would also like to emphasize that the low energy excitation is only observable with momentum transfer along the ΓF direction (or weakly along ΓM) and not along the ΓK direction, consistent with the notion that K point is not a time-reversal invariant momentum point in the 3D Brillouin zone of $Bi_2Se_3$.[35]

## IV. CONCLUSION

The electronic excitations including the plasmons and interband transitions in single-crystal $Bi_2Se_3$ were thoroughly investigated by electron energy-loss spectroscopy in conjunction with (scanning) transmission electron microscopy, (S)TEM-EELS. Both the volume plasmons (at 7 and 17 eV) and surface plasmons (at 5.5 and 10 eV) are demonstrated with positions-dependent STEM-EELS spectra and real-space intensity mapping in STEM-EELS spectral imaging. The momentum-dependent EELS was employed to investigate π and π+σ volume plasmons dispersion with momentum transfer ($q$) parallel to in-plane (ΓM and ΓK) and out-of-plane (ΓZ) directions, and no significant anisotropy was found. However, the π plasmon exhibited a linear dispersion as a function of $q$, in analogy to graphene, whereas the π+σ plasmon displayed a conventional parabolic dispersion. The physical origin for the linear dispersion of the π plasmon is not clear at the present time and more work will be needed to clarify this issue. Furthermore, a low energy excitation 0.7~1.6 eV was also observed which is attributed to direct nonvertical interband transitions along the ΓF direction. The dispersion of this excitation was found follows closely to the shape of the lowest conduction band along the ΓF direction.

This work was supported by the National Science Council of Taiwan (NSC99-2112-M-002-027-MY3 and NSC101-2811-M-002-130).

**FIGURE CAPTIONS**

FIG. 1. (Color online) (a) STEM–EELS spectra acquired on single-crystal $Bi_2Se_3$ with the electron probe positioned at various locations of the material (inset HAADF image; probe step, 2 nm). The color circles (inset) denote probe locations and the corresponding EELS spectra are shown by the same colors. The intensities of the EELS spectra were normalized to ZLP and then ZLP was removed. Solid black arrow indicates the spectrum acquired at the sample edge. (b) The complex dielectric function of $Bi_2Se_3$ derived from the blue spectrum in (a). The complex dielectric function below 2.5 eV was reproduced from Ref. 23. The pink, sky blue, blue and green spectra are the blow-ups of those in (a) with removed ZLP for clarity. Inset is the calculation of effective number $n_{eff}(\omega)$ of electrons per atom in $Bi_2Se_3$. (c-e) the STEM-EELS excitation-intensity maps acquired from the HAADF image (inset, (a)) at different energy: (c) ~ 5.5 eV, (d) 10 eV, and (e) 17 eV. The dashed black lines in (c-e) indicate the interface between material (upper part) and vacuum (bottom part). Color scale bar represents the linearly normalized image intensity.

FIG. 2. Momentum-dependent EELS spectra obtained along the momentum transfer ($q$) parallel to the (a) ΓM, (b) ΓK and (c) ΓZ directions. The spectra are all normalized to the intensity of peak at 17 eV and then displaced vertically to improve readability. Inset in (a-c) are the selected-area electron diffraction (SAED) patterns of $Bi_2Se_3$ acquired with the incident electron beam parallel to the c-axis along [001] direction (inset, (a) and (b)), and perpendicular to the c-axis along [$\bar{1}$10] direction (inset, (c)). The (hexagonal) dash lines in (a) and (b) illustrates the Brillouin zone. The arrows with white/red color in (a-b) and (c) indicate the momentum transfer range for momentum-dependent EELS experiments.

FIG. 3. (color online) (a) Lorentz curve fitting of EELS spectra for $q$= 0, 0.8 and 1.28 Å$^{-1}$, respectively. (b) The energy of $\pi$ and $\pi+\sigma$ plasmons along ΓM, ΓK and ΓZ directions as a function as momentum transfer ($q$) with black square (ΓM direction), gray circle (ΓK direction) and red triangle (ΓZ direction), respectively.

FIG. 4. (color online) (a) Momentum-dependent EELS spectra of additional peak at 0.7~1.6 eV observed with sample tilting. (b) The dispersion curve of the nonvertical

intenband transition peaks as a function of momentum-transfer with sample tilting (marked with black square, respectively). The dashed line is the linear curve fitting. The dispersion of the lowest conduction band along ΓF direction in ref. 12 with energy normalized to the experimental value of energy gap ($E_g$= 0.32 eV) at $q$= 0 is also plotted for comparison.

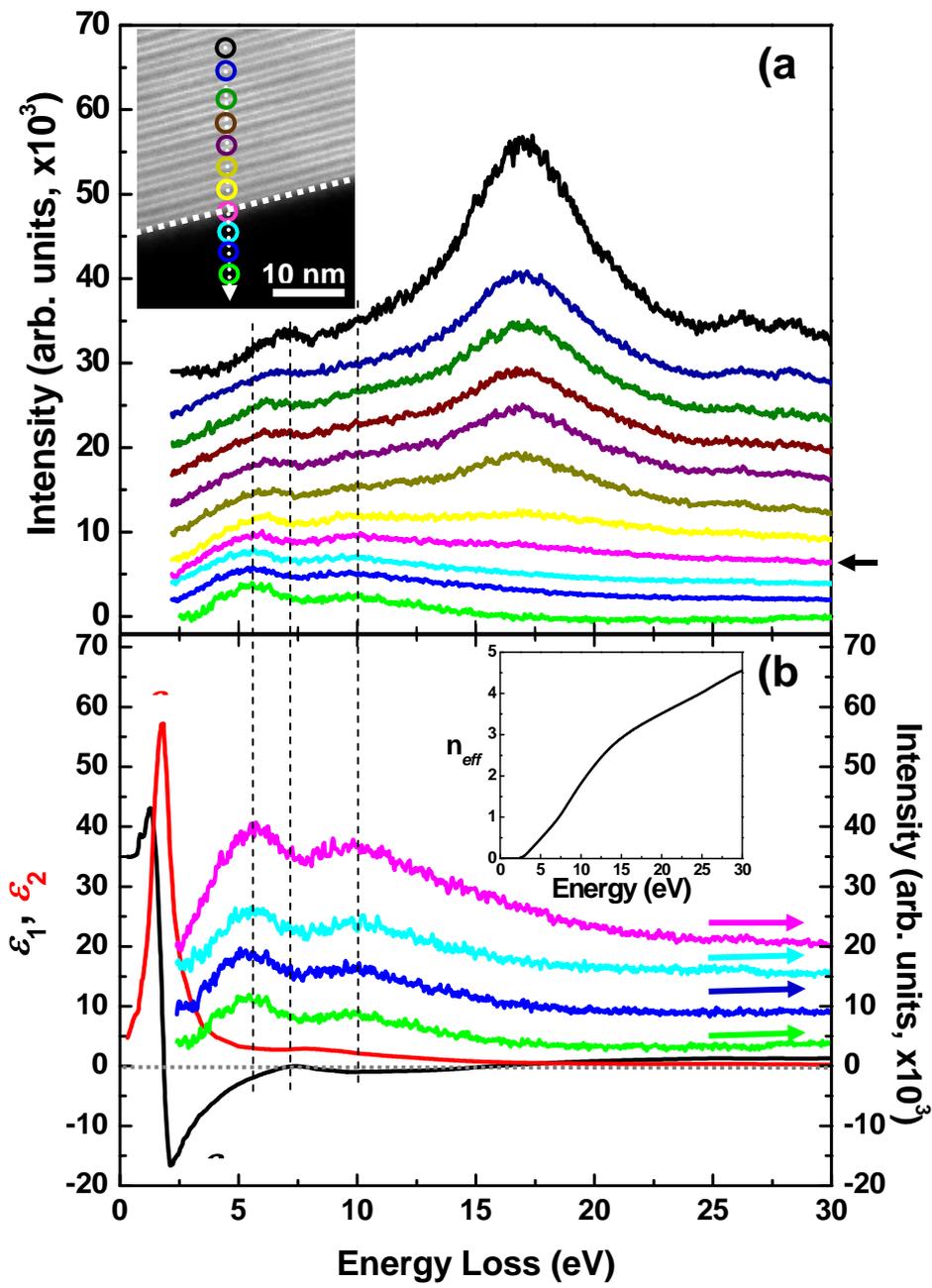

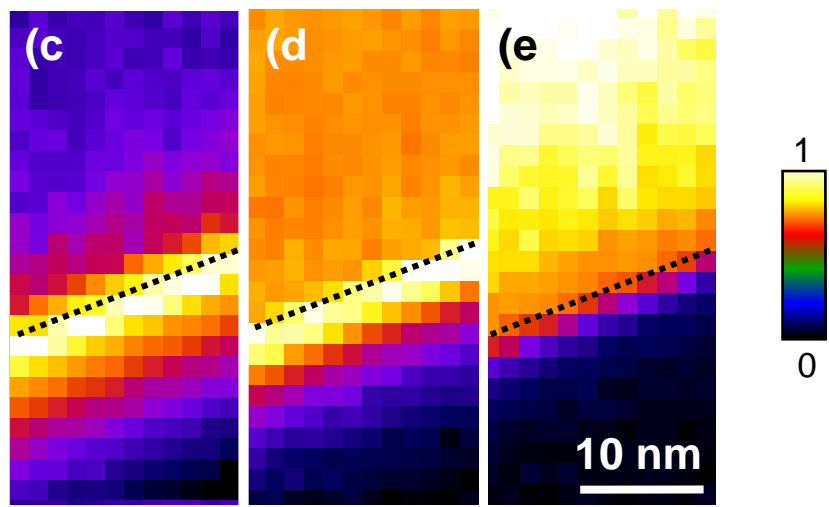

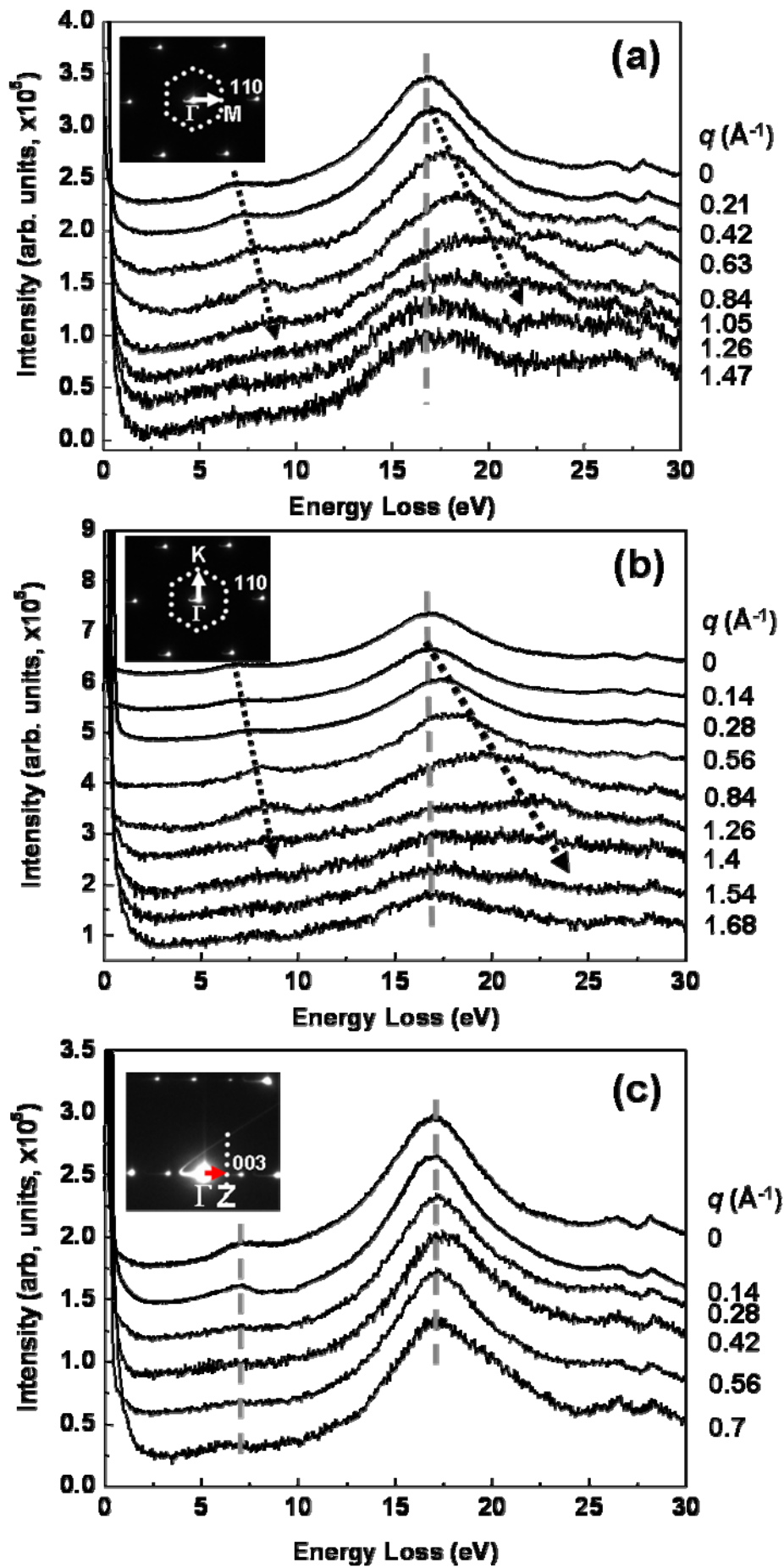

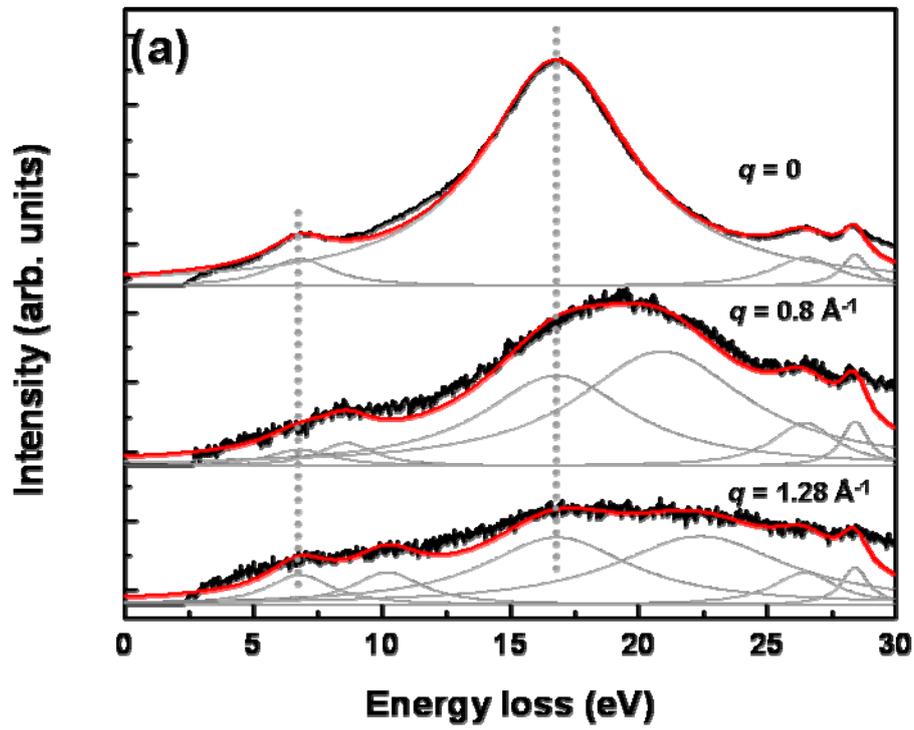

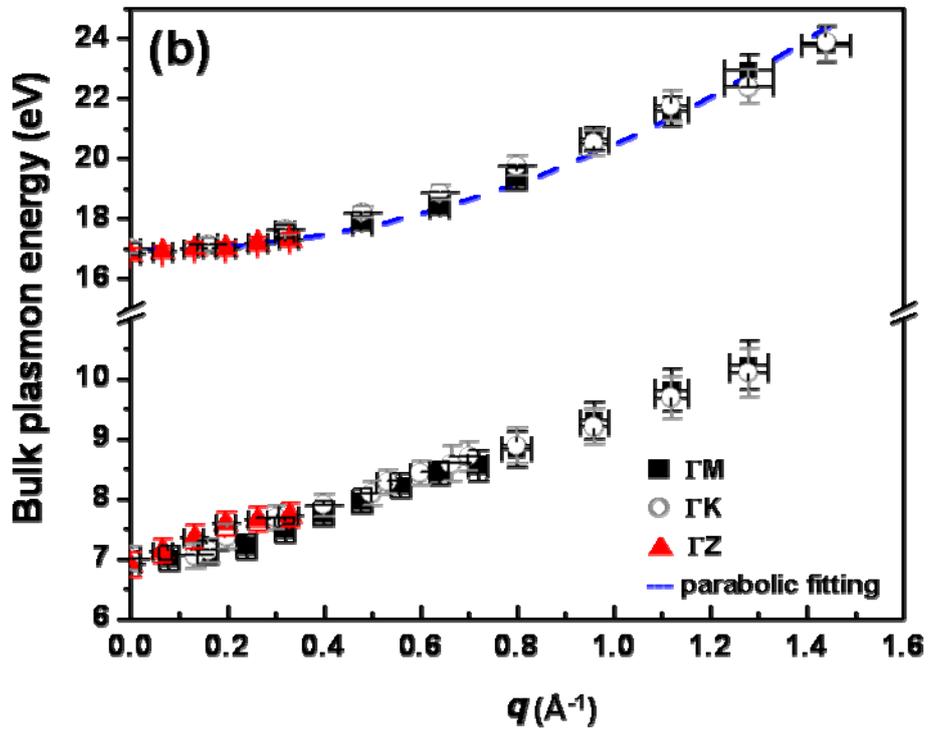

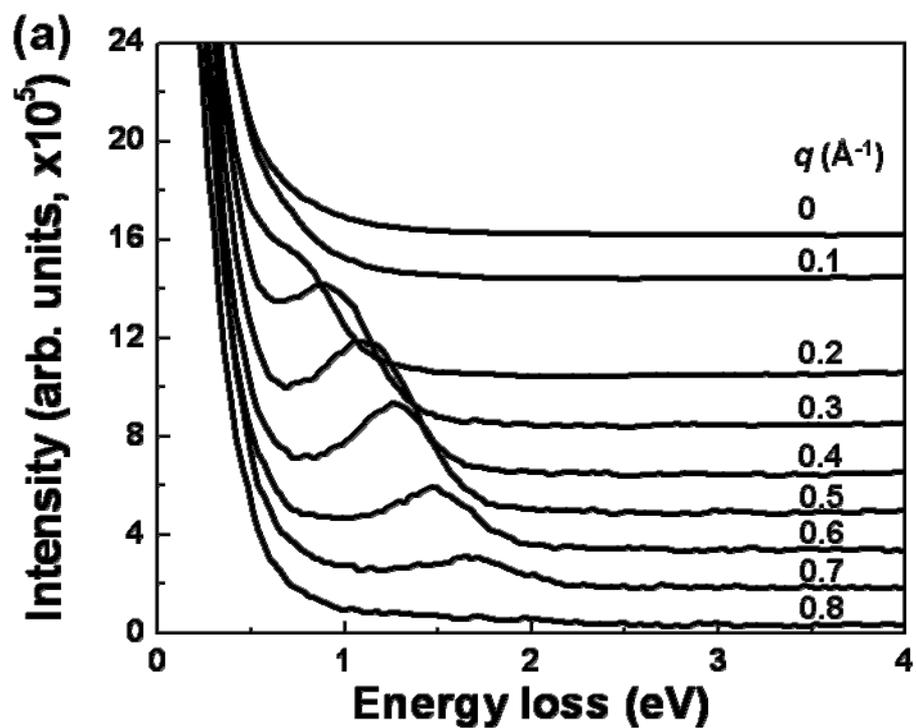

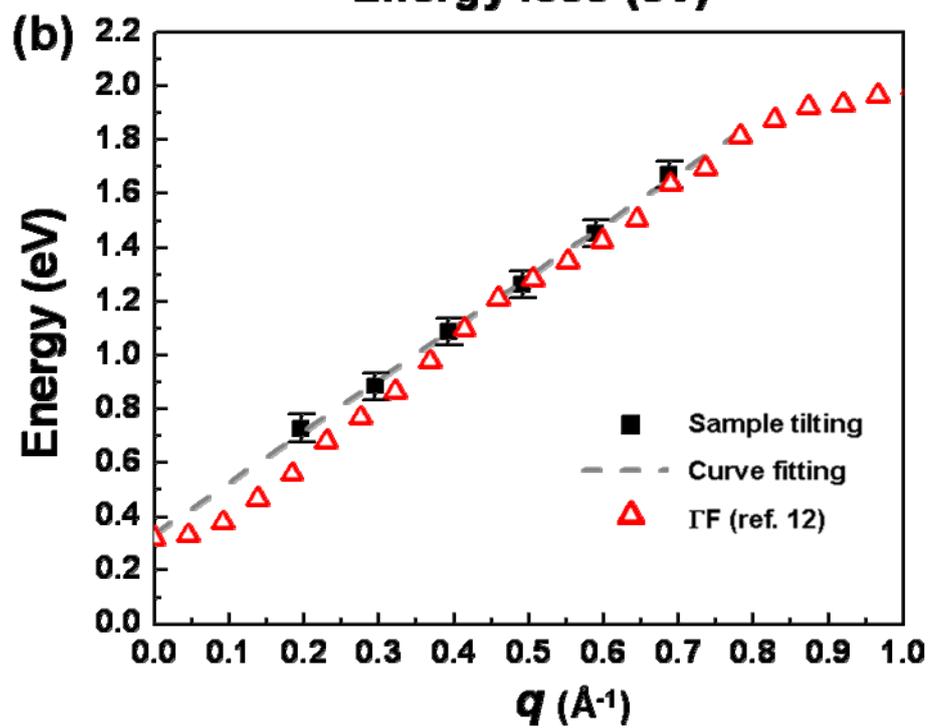